\documentclass[fleqn]{eptcs}
\usepackage{amssymb}
\usepackage{amsmath}
\usepackage{graphicx}
\usepackage{stmaryrd}

\title{A Process Algebra for Supervisory Coordination}
\author{Jos Baeten \qquad Bert van Beek \qquad Allan van Hulst \qquad Jasen Markovski\thanks{Research funded by C4C European project FP7-ICT-2007.3.7.c.}
\institute{
Department of Mechanical Engineering,
Eindhoven University of Technology,\\
P.O.~Box~513, 5600 MB Eindhoven, The Netherlands,}
\email{\{j.c.m.baeten,d.a.v.beek,ahulst,j.markovski\}@tue.nl}
}
\def\titlerunning{A Process Algebra for Supervisory Coordination}
\def\authorrunning{Baeten, van Beek, van Hulst, Markovski}

\renewcommand{\mathcal}{\mathsf}

\newtheorem{definition}{Definition}

\renewcommand{\supset}{\Rightarrow}
\newcommand{\bbbn}{\mathbb{N}}

\newcommand{\TCPstar}{\ensuremath{\mathrm{TCP}^{*}}}

\newcommand{\ActC}{\ensuremath{\mathalpha{\mathcal{A}}}}
\newcommand{\act}[1][a]{\ensuremath{\mathalpha{#1}}}

\newcommand{\Data}{\ensuremath{\mathcal{D}}}
  
\newcommand{\datum}[1][d]{\ensuremath{\mathalpha{#1}}}

\newcommand{\emptystr}{\ensuremath{\mathalpha{\varepsilon}}}
\newcommand{\Chan}{\ensuremath{\mathcal{H}}}
\newcommand{\chan}[1][c]{\ensuremath{\mathalpha{#1}}}
  \newcommand{\rec}[2][\chan]{\ensuremath{#1?{#2}}}
  \newcommand{\snd}[2][\chan]{\ensuremath{#1!{#2}}}
  \newcommand{\com}[2][\chan]{\ensuremath{#1!\!?{#2}}}

\newcommand{\PEXP}[1][T]{\ensuremath{\mathalpha{\mathcal{#1}}}}
\newcommand{\States}[1][P]{\ensuremath{\mathalpha{\mathcal{P}}}}
  
\newcommand{\state}[1][s]{\ensuremath{\mathalpha{#1}}}
  \newcommand{\states}[1][]{\ensuremath{\state[s_{#1}]}}
  \newcommand{\statet}[1][]{\ensuremath{\state[t_{#1}]}}
  
\newcommand{\trel}[1][]{\ensuremath{\mathalpha{\longrightarrow_{#1}}}}

\newcommand{\final}[1][]{\ensuremath{\mathalpha{\downarrow_{#1}}}}

\newdimen\boxwdplusemdimen
  \def\arrow#1{{
     \boxwdplusemdimen=1em%
     \setbox0=\hbox{$\scriptstyle#1$}%
     \advance\boxwdplusemdimen by \wd0\relax%
     \ifdim\boxwdplusemdimen<16.11119pt%
       \boxwdplusemdimen=16.11119pt%
     \fi%
     \buildrel{#1}\over%
       {\setbox1=\hbox to \boxwdplusemdimen{\rightarrowfill}%
     \ht1=0.3em\relax\box1}%
   }}
\newcommand{\step}[1]{\mathbin{\stackrel{#1}{\mathord{\longrightarrow}}}}
\newcommand{\nostep}[1]{\mathbin{\stackrel{#1}{\mathord{\mkern5mu \arrownot \mkern-5mu \longrightarrow}}}}
\newcommand{\steps}[1]{\mathbin{\stackrel{#1}{\mathord{\twoheadrightarrow}}}}

\newcommand{\term}[2][]{\ensuremath{{#2}\mathclose{\downarrow_{#1}}}}

\newcommand{\lang}[1]{\ensuremath{\mathcal{L}(#1)}}

\newcommand{\dl}{0}
\newcommand{\emp}{1}
\renewcommand{\merge}{\parallel}
\newcommand{\encap}[2]{\partial_{#1}(#2)}
\newcommand{\Act}{\mathcal{A}}
\newcommand{\rse}[2]
  {\ensuremath{{#1}\mkern 1.5mu \rlap{\raisebox{.5ex}{$\scriptstyle \wedge$}}\mkern 4.5mu \raisebox{.5ex}{$\scriptscriptstyle \blacktriangle$}{#2}}}

\newcounter{soscounter}

\newcommand{\osrule}[2]{
    \refstepcounter{soscounter}
    \mbox{\small $\mathbf{\thesoscounter}~$}
    \displaystyle \frac{#1}{#2}
}

\begin{document}


\maketitle
\begin{abstract}
A supervisory controller controls and coordinates the behavior of different components of a complex machine by observing their discrete behaviour. Supervisory control theory studies automated synthesis of controller models, known as supervisors, based on formal models of the machine components and a formalization of the requirements. Subsequently, code generation can be used to implement this supervisor in software, on a PLC, or embedded microprocessor.
In this article, we take a closer look at the control loop that couples the supervisory controller and the machine. We model both event-based and state-based observations using process algebra and bisimulation-based semantics. The main application area of supervisory control that we consider is coordination, referred to as supervisory coordination, and we give an academic and an industrial example, discussing the process-theoretic concepts employed.
\end{abstract}

\section{Introduction}

\begin{figure}[!h]
\centering
\includegraphics[width = \linewidth]{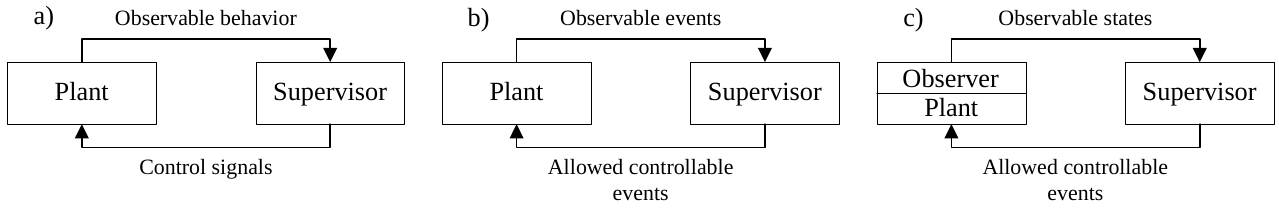}
\caption{Control loop: a) general, b) with event-based, c) with state-based observations.}\label{fig:control-loop}
\end{figure}

Control software development becomes an important issue due to the ever-increasing machine complexity and demands for better quality, performance, safety, and ease of use. Traditionally, the control requirements are formulated informally and, thereafter, manually translated into control software, followed by validation and rewriting of the code whenever necessary. The cycles of such a design-validate process are both error-prone and time-consuming due to frequent ambiguities in the informal specifications. This issue gave rise to supervisory control theory~\cite{rwsupervisor,Cassandras,MaWonham}, where models of the supervisory controllers, referred to as \emph{supervisors} are synthesized automatically based on formal models of the uncontrolled hardware, referred to as \emph{plant}, and the model of the \emph{control requirements}. Based on these models, the control software is generated automatically. The supervisory controller observes discrete machine behavior and sends back control signals about allowed activities. Assuming that the controller reacts sufficiently fast on machine input, this feedback loop, depicted in Figure~\ref{fig:control-loop}a), was originally modeled as a pair of synchronizing processes~\cite{rwsupervisor,Cassandras}.

In this paper, we focus on the modeling of the control loop and the required process-theoretic concepts to capture the underlying behavior. The main motivation for the investigation is the oversimplification of the coupling between the plant and the supervisor in the original proposal of~\cite{rwsupervisor,Cassandras}, which still prevails in modern state-of-the-art approaches, like~\cite{heymann-nondeterministic,fabiannondeterministic,overkamp-nondeterministic,kumarnondeterministic,kumar-simulation} to name a few. Furthermore, we consider coordination as the main application area of supervisory control, where the coordinator(s) are implemented as supervisory controllers that ensure sequencing of events, or deadlock- and livelock-free behavior of the plant, according to the given set of (coordinating) control requirements.

\paragraph{Supervisory control loop} To model different aspects of the plant and the control requirements, the discrete events that can occur are split into \emph{controllable} and \emph{uncontrollable} events. The former can be disabled by the supervisor and typically model actuator activities, e.g., starting or stopping a motor. The latter cannot be affected by the supervisor if enabled in the plant and standardly model sensor activities, e.g., the temperature has reached a given value. We distinguish two types of prominent supervisory control loops relying on event- and state-based observations, depicted in Figure~\ref{fig:control-loop}b) and c), respectively.

The control loop in Figure~\ref{fig:control-loop}b) depicts that the supervisor observes events that occur in the plant and sends back as feedback the set of controllable events that are allowed for execution. The most prominent operation that captures the coupling between the plant and the supervisor is automata-style synchronous parallel composition~\cite{rwsupervisor,Cassandras}. This simple operation restricts the plant by omitting (controllable) events in the supervisor, thereby preventing synchronization and disabling the events. It was quickly realized that this synchronization produces large supervisors that actually memorize the complete supervised behavior as the supervisor keeps track of the state of the plant by keeping a complete history of observed events.

To mitigate the large size of the supervisor several synchronization operators were proposed that enable the plant to independently execute uncontrollable events, provided that this does not preclude the supervisor from correctly deducing the state of the plant~\cite{nasa-algebra,heymann-nondeterministic,kumar-prioritized,PCXG}. We note that there are other models of the control loop that employ the input/output transition paradigm that require an input (set of controllable/actuator events) from the supervisor to produce an output (uncontrollable/sensor event) from the plant~\cite{iosupervisor,lafortunebisimulation,iokumar}. Nonetheless, they have been shown to be equivalent to one of the above approaches with respect to the underlying notions~\cite{lafortunebisimulation}.

What synchronous parallel composition or communication fails to model is the difference between the two flows of information, their role, and the different goals of the plant and the supervisor. To this end, we propose a send/receive communication to model the different flows of communication in Figure~\ref{fig:control-loop} and differentiate between the contributions of the plant and the supervisor. The event-based observation flow of Figure~\ref{fig:control-loop}b) enables communication of all observable events, whereas the control signal flow transmits only controllable events. In addition, this setting also supports asynchronous communication between the plant and the supervisor, which affects almost every implementation of supervisory controllers~\cite{asynchronous}.

As a solution to the problem of large supervisors, an alternative approach was proposed in~\cite{MaWonham}, as depicted in Figure~\ref{fig:control-loop}c). The plant (or the supervisor) is augmented with an observer or a tracker that deduces the state of the plant and submits this observed information to the supervisor. The supervisor, based on this state-based information, acts as a lookup table and feeds the plant with the allowed controllable events in the observed state. In such a way, the supervisor only incorporates necessary information in order to exercise control over the plant. Nonetheless, this feedback mechanism is not formalized in~\cite{MaWonham} and, here, we propose to model this variant of the control loop using a process-theoretic approach that employs root signal emission~\cite{pabook} to capture the state-based observations. Alternative modeling of such control loops is by means of shared variables and synchronization~\cite{extendedguards}, but such approaches do not distinguish between the different flows of information depicted in Figure~\ref{fig:control-loop}c).

Finally, when employing supervisory control for coordination of distributed systems, the supervisor communicates the control actions to several components that have different physical locations. To this end, we propose to model the feedback control signal communication from the supervisor by means of broadcast communication~\cite{pabookold}. To illustrate the proposed process theories that capture the behavior of the control loops of Figure~\ref{fig:control-loop}b) and c) we revisit two cases, where we applied supervisory coordination: \ref{fig:control-loop}b) a simple case that introduces the main concepts and deals with coordination of an automated guided vehicle, involving event-based observations, and \ref{fig:control-loop}c) a part of an industrial study dealing with maintenance procedures inside complex high-tech printers, which employs state-based observations~\cite{wodes2010}.

\paragraph{Process-theoretic approach} The process-theoretic treatment of supervisory control theory is sustained by a behavioral relation that captures the notion of \emph{controllability}, which states that supervisory control is possible only if the supervisor can achieve supervised behavior allowed by the control requirements without having to disable an uncontrollable event. Prior investigations to process-theoretic treatments of supervisory control resulted in a special prioritized synchronization operator~\cite{nasa-algebra,heymann-nondeterministic}, while employing failure semantics. An alternative approach replaced this special operator with a refinement relation to characterize nondeterministic supervised behaviors~\cite{overkamp-nondeterministic}. In \cite{kumarnondeterministic,tabuadabisimulation} the refinement is given in terms of bisimulation and in terms of simulation in~\cite{kumar-simulation}. A coalgebraic approach introduced partial bisimulation as a behavioral relation suitable to define language-based controllability~\cite{coalgebra}. In essence, it states that controllable events should be simulated, whereas uncontrollable events should be bisimulated. This notion was lifted to a concurrency theory for supervisory control that succinctly captured the controllability for nondeterministic discrete-event systems~\cite{acc2011}. Here, we extend this framework to elaborately model and formalize the behavior of the supervisory control loops depicted in Figure~\ref{fig:control-loop}.

The rest of this paper is organized as follows. Section 2 revisits the process theory $\TCPstar$ from~\cite{pabook} and establishes a link between partial bisimulation and supervisory control. Section 3 shows how to model supervisory control loop in the presented theory by applying supervisory coordination to an automated production line. Section 4 extends the process theory to incorporate guarded commands and root signal emission, which are employed in Section 5, where we revisit an industrial case study of coordination of maintenance procedures in a high-tech printer. We finish with a discussion of future challenges and the potential of applying process theory in supervisory control.

\newcommand{\comn}[4]{\ensuremath{#1!_{#2}\!?_{#3}{#4}}}
\newcommand{\pbis}[1]{\mathbin{\leq_{#1}}}
\newcommand{\bpbis}[1]{\mathbin{\leftrightarrow_{#1}}}
\newcommand{\Uact}{\mathcal{U}}
\newcommand{\Cact}{\mathcal{C}}
\newcommand{\supp}[2]{\ensuremath{#1 / #2}}

\section{Process theory $\TCPstar$}\label{sec:TCP}

In this section we revisit the process algebra \TCPstar{} (Theory of Communicating Processes with Iteration)~\cite{pabook} in which we introduce \emph{generic communication actions} and we adopt \emph{partial bisimulation} as a behavioral relation. This process algebra has a rich syntax, allowing us to express all key ingredients of concurrency theory, including termination, which enables a strong correspondence with automata theory.

\paragraph{Syntax}

We presuppose a finite \emph{data alphabet} $\Data$ and a finite set $\Chan$ of \emph{channels}. We assume that $\ActC  = \{ \comn{\chan}{m}{n}{\datum} \mid \chan\in\Chan, \ m, n \in \bbbn, \ \datum\in\Data\}$, where $\comn{\chan}{m}{n}{\datum}$ is a generic communication action. If $m = n = 0$, then we treat the generic communication action $\comn{c}{0}{0}{d}$ as a basic event, possibly parameterized with data, notation $c(d)$. Otherwise, we handle it as an outcome from synchronization of $m$ send and $n$ receive actions. We employ the standard notation for handshaking communication~\cite{pabook}, i.e., $\rec{d}$ for $\comn{\chan}{0}{1}{\datum}$, $\snd{d}$ for $\comn{\chan}{1}{0}{\datum}$, and $\com{\datum}$ for $\comn{\chan}{1}{1}{\datum}$. Intuitively, these events denote that datum $\datum$ is received, sent, or communicated along channel $\chan{}$, respectively.

The set of \emph{process terms} $\PEXP$ is generated by the following grammar:
  \[
    T ::=
      \dl\ \mid\ \emp\ \mid\
      a.T\ \mid\
      T \cdot T \ \mid\ \ T + T\ \mid\
      T \merge T\ \mid\ T^{*}\ \mid\
      \encap{E}{T},
\]
where  $a\in\ActC$ and $E \subseteq \{ \comn{\chan}{m}{n}{} \mid \chan\in\Chan, \ m, n \in \bbbn\}$. Let us briefly comment on the operators in this syntax. The constant $\dl$ denotes inaction or \emph{deadlock}, whereas the constant $\emp$ denotes \emph{successful termination}~\cite{pabook}. For each action $\act\in\ActC$ there is a unary operator $a.{}$ denoting \emph{action prefix}; the process denoted by $a.{p}$ can do an $\act$-transition to the process denoted by $p$.
The binary operator $p\cdot q$ denotes \emph{sequential composition} that behaves like $p$, followed by $q$ only upon successful termination of $p$. The binary operator $p+q$ denotes \emph{alternative composition} or choice on the first action transition of $p$ and $q$. The binary operator ${p \merge q}$ denotes \emph{parallel composition (with generic channel communication actions)}; actions of both arguments can be interleaved or, alternatively, communication takes place that keeps track of how many send or receive actions are combined. The unary operator $p^{*}$ is \emph{iteration} or Kleene star that unfolds with respect to the sequential composition. The unary operator $\encap{E}{p}$ \emph{encapsulates} the process $p$ in such a way that all (incomplete) communication actions, e.g., $\rec[\chan]{\datum}$ and $\snd[\chan]{\datum}$, are blocked for all data, so that the desired type of communication is enforced, e.g., if we were to enforce communication between $k$ processes on channel $c$, then $E = \{\comn{c}{m}{n}{} \mid 0 < m + n, \ m + n \neq k\}$.


\paragraph{Semantics}

We give semantics to the process terms by a labeled transition relation $\mathalpha{\step{}} \subseteq\PEXP\times\Act\times\PEXP$ and a successful termination predicate $\mathalpha{\final}\in\PEXP$.
%
We employ infix notation and write $\states \step{a} \statet$ if $(\states,a,\statet) \in \trel$ and $\term{\states}$ if $\states \in \final$.
%
%
We derive the transition relation and the successful termination predicate using structural operational semantics~\cite{Plo04a}, given by the operational rules in Table~\ref{tab:TCPsostable}.
Alternatively, we depict them as a labeled transition system $G$, specified by the tuple $G = (\PEXP, \ActC, \final{}, \step{})$.

\begin{table}[t!]
\centering
\begin{tabular}{c}
\[\hspace{-0.5cm}
\begin{array}{c}
   \osrule{}{\emp \downarrow}
 \qquad
   \osrule{}{a.{p} \step{a} p}  \qquad
 \osrule{}{p^{*} \downarrow} \qquad
 \osrule{p \step{a{}} p'}%
   {p^{*} \step{a{}} p' \cdot p^{*}}
\qquad   \osrule{p \downarrow, \ q \downarrow}%
           {p \cdot q \downarrow} \qquad
   \osrule{p \downarrow, \ q \step{a{}} q'}%
           {p \cdot q \step{a{}} q'} \smallskip \\
   \osrule{p \step{a{}} p'}%
           {p \cdot q \step{a{}} p' \cdot q}
\qquad   \osrule{p \downarrow}{(p + q) \downarrow} \qquad
   \osrule{q \downarrow}{(p + q) \downarrow}\qquad
   \osrule{p \step{a{}} p'}%
           {(p + q) \step{a{}} p'} \qquad 
   \osrule{q \step{a{}} q'}%
           {(p + q) \step{a{}} q'}
\smallskip \\

   \osrule{p \downarrow, \  q \downarrow}%
           {p \merge q \downarrow}
           \qquad
   \osrule{p \step{a} p'
   }%
           {p \merge q \step{a{}} p' \merge q}
 \qquad
   \osrule{q \step{a{}} q'
    }%
           {p \merge q \step{a{}} p \merge q'}
\qquad
   \osrule{p \downarrow}{\encap{E}{p} \downarrow}
\smallskip \\
   \osrule{p \step{\comn{\chan}{\ell}{k}{\datum}} p', \
            q \step{\comn{\chan}{m}{n}{\datum}} q', \ \ell + k > 0, \ m + n > 0 }%
           {p \merge q \step{\comn{\chan}{\ell + m}{k + n}{\datum}} p' \merge
             q'}
\qquad
      \osrule{p \step{a{}} p',
              \ a \not \in \{\comn{c}{m}{n}{d} \mid \comn{c}{m}{n}{} \in E, \ d \in \Data\}}%
           {\encap{E}{p} \step{a{}}
             \encap{E}{p'}}
\smallskip \\
\end{array}
\]
\end{tabular}
 \caption{Operational rules for \TCPstar{}%
}\label{tab:TCPsostable} %
\end{table}

We briefly comment on the rules. The successful termination constant can successfully terminate, whereas the action prefix enables outgoing labeled transitions, as given by rules 1 and 2. Rule 3 states that iteration can always terminate successfully, which enables sequential composition of recursive processes. The unfolding of the iteration is with respect to the sequential composition, as given by rule 4. The sequential composition can terminate only if both processes can do so, as given by rules 5, whereas if only the first component can terminate successfully, it can continue behaving as the second. The outgoing transition of the first component is the same for the sequential composition as given by rule 7. Rules 8 and 9 state that alternative composition can terminate if one of the components can terminate, whereas the choice is made on the outgoing transitions, as stated by rules 10 and 11. The parallel composition can terminate only if both components can do so. Rules 13 and 14 enable interleaving of transitions. Rules 15 states that encapsulation does not prevent successful termination. Rule 16 defines synchronization which can occur on communication actions comprising at least one sending or receiving event. The communication actions are merged to accumulate the participating send and receive parties. Finally, rule 17 states that all (incomplete) communication actions on a given channel comprising a predefined number of senders and receivers are blocked by the encapsulation operation.

We can easily extend the transition relation to traces of actions in $\ActC^*$. For $p, p' \in \PEXP$ and $t = a_1,\dots,a_n \in \ActC^{*}$, we write $p \steps{t} p'$ if there exist $p_0,\dots,p_n\in\PEXP$ such that $p=p_0 \step{a_1}\cdots\step{a_n} p_n=p'$. By $\emptystr$ we denote the empty trace $a_1,\dots,a_n$ for $n = 0$ and $p = p'$. Every finite automaton can be described up to isomorphism (and possibly by changing the communication operation) by a term in our setting, see \cite{BLMT10}.

\paragraph{Language-based supervision}

Now, we can translate the central notion of a supervisor~\cite{rwsupervisor,Cassandras} in our setting. To this end, we partition the channel names into two disjoint sets of uncontrollable~$\Uact$ and controllable~$\Cact$ channels such that $\Chan = \Uact \cup \Cact$ and $\Uact \cap \Cact = \emptyset$. The uncontrollable and controllable channel names induce controllable and uncontrollable actions, respectively, given by $\Act_\Cact \triangleq \{\comn{\chan}{m}{n}{\datum} \mid c \in \Cact, \ d \in \Data\}$ and $\Act_\Uact \triangleq \{\comn{u}{m}{n}{\datum} \mid u \in \Uact,\ d \in \Data\}$. Next, we define the (prefix-closed) language recognized by the process term $p$ or, alternatively, the automaton represented by $p$, as $\lang{p} \triangleq \{ t \in \ActC^* \mid \mathrm{there \ exists\ } p' \in \PEXP \mathrm{\ such \ that \ } p \steps{t} p'\}$. Note that traces do not need to end with successful termination. We denote by $L L' \triangleq \{t t' \mid t \in L, \ t' \in L'\}$ the concatenation of the languages $L$ and $L'$.

Recall that the supervisor cannot achieve the control requirements by forbidding uncontrollable events, when synchronizing with the plant. Suppose that the plant, the control requirements, and the supervisor with respect to the former are determined by the languages recognized by the process terms $p, r, s \in \PEXP$, respectively. If the operation modeling the control loop is denoted by $\supp{p}{s}$, then $\lang{\supp{p}{s}} \subseteq \lang{p}$ and $\lang{\supp{p}{s}} \subseteq \lang{r}$, where we refer to $\supp{p}{s}$ as the \emph{supervised plant}. We note that if strict equality holds, then the control requirements can be achieved completely. Often, this is not the case, so one attempts to synthesize a maximally-permissive supervisor, which makes $\lang{\supp{p}{s}}$ as large as possible with respect to inclusion. For deterministic systems, this supervisor is unique, equal to the union of all possible supervisors~\cite{rwsupervisor,Cassandras}, whereas for nondeterministic systems, a unique maximally-permissive supervisor in general does not exist~\cite{acc2011}. For standard supervisory control~\cite{rwsupervisor,Cassandras}, the operation that models the control loop $\supp{p}{s}$ is the full synchronous parallel composition of automata~\cite{rwsupervisor,Cassandras}. That $s$ does not disable uncontrollable events is ensured by requesting that $\supp{p}{s}$ is \emph{controllable} with respect to $p$, expressed as $\lang{\supp{p}{s}} \Uact \cap \lang{p} \subseteq \lang{\supp{p}{s}}$~\cite{rwsupervisor,Cassandras}. Controllability is interpreted as follows. If we observe a desired trace in the plant followed by an uncontrollable event, then the supervisor cannot request that this uncontrollable event should be disabled after allowing that trace. If $r$ is controllable with respect to $p$, then one can guarantee the existence of a supervisor $s$, achieving the desired controlled behavior $r$ by restricting the plant $p$ by synchronization, i.e., $\lang{\supp{p}{s}} = \lang{r}$.

\paragraph{Nondeterminism and partial bisimulation}

The disadvantages of working in the language domain have been discussed on many occasions, e.g., see overviews in~\cite{heymann-nondeterministic,fabiannondeterministic,acc2011,pabook}. Therefore, a proposal was made in~\cite{acc2011} to lift controllability to support full nondeterminism in a process-theoretic setting. The underlying behavioral relation is partial bisimulation~\cite{coalgebra,acc2011}, which is parameterized with a bisimulation actions set $B \subseteq \ActC$ that denotes which actions are to be bisimulated, whereas the other actions are simulated.

\begin{definition}\label{def:partial-bisimilarity}
A relation~$R \subseteq \PEXP \times \PEXP$ is a partial bisimulation with respect to a bisimulation action set~$B \subseteq \ActC$, if for all $(p,q) \in R$ it holds that:
\begin{enumerate}
\item  if $p \downarrow$, then $q \downarrow$;

\item 
if $p \step{a} p'$ for some $a \in \ActC$, then there exists $q' \in \PEXP$ such that $q \step{a} q'$ and $(p',q') \in R$;

\item 
if $q \step{b} q'$ for some $b \in B$, then there exists $p' \in \PEXP$ such that $p \step{b} p'$ and $(p',q') \in R$.
\end{enumerate}
If $(p, q) \in R$, we say that $p$ is partially bisimilar to $q$ with respect to $B$ and we write $p \pbis{B} q$. If $q \pbis{B} p$ holds as well, we write $p \bpbis{B} q$.
\end{definition}
Note that $\pbis{B}$ is a preorder relation, making $\bpbis{B}$ an equivalence relation for all $B\subseteq \ActC$~\cite{acc2011}. If $B=\emptyset$, then $\pbis{\emptyset}$ coincides with strong similarity preorder and $\bpbis{\emptyset}$ coincides with strong similarity equivalence~\cite{glabbeek,pabook}. When $B = \ActC$, $\bpbis{\ActC}$ turns into strong bisimilarity~\cite{glabbeek,pabook}. Moreover, if $p \mathbin{\pbis{B}} q$, then $p \mathbin{\pbis{C}} q$ for every $C \subseteq B$. We also note that partial bisimilarity is a precongruence with respect to the operators of $\TCPstar$~\cite{acc2011}.

For given processes $p, r \in \PEXP$, representing the plant and the control requirements, respectively, we ensure that $s \in \PEXP$ is a valid supervisor that does not disable uncontrollable events by requiring that $\supp{p}{s} \pbis{\emptyset} r$ and $\supp{p}{s} \pbis{\ActC_\Uact} p$, where $\ActC_\Uact \subseteq \ActC$ is the set of uncontrollable events~\cite{acc2011}. This setting covers both the existing deterministic and nondeterministic definition of controllability for discrete-event systems~\cite{acc2011}.
From the definition, it is also not difficult to observe, that one obtains the same supervised behavior for every $p'\bpbis{\ActC_\Uact} p$. Thus, one direct benefit from our approach is a procedure for coarsest plant minimization that respects controllability, based on the partial bisimilarity equivalence. 

Next, we model the supervisory control loop with event-based observations and we illustrate our approach by a use case involving coordination of an automated guided vehicle in a production line.

\newcommand{\ren}{\xi}

\section{Control Loop with Event-Based Observations}\label{sec:regproc}

We employ the process theory $\TCPstar$ to formalize the behavior of the control loop with event-based observations, depicted in Figure~\ref{fig:control-loop}b). According to the scheme, the plant cannot execute a controllable event without the permission of the supervisor, whereas the supervisor must not disable uncontrollable events. Nonetheless, the supervisor is able to observe execution of uncontrollable events in the plant, so that it can correctly determine the state of the plant and transmit correct control signals. Moreover, the supervisor should not execute uncontrollable events independently, as this does not contribute to his objective. In addition, the supervisor should not introduce deadlocks or livelocks explicitly, unless deadlock or livelock behavior is inherent to the plant. Finally, we assume that the supervisor is a (global) monolithic process, i.e., it is not comprised from multiple modular or distributed synchronizing supervisors~\cite{Cassandras}. Taking into account the above observations, we can specify the syntax of the plant processes~$\mathcal{P}$ and the supervisor processes~$\mathcal{S}$ as given by $P$ and $S$, respectively:
\[
\begin{array}{l}
P ::= \dl\,\mid\,\emp\,\mid\, c?d.P\,\mid\, \comn{u}{\ell}{k}{d}.P\,\mid\, P \cdot P \, \mid\,\,P + P\,\mid\, P \merge P\,\mid\,\encap{E}{P}\,\mid\,P^{*}
\smallskip \\
S ::= \emp\ \mid c!d.S\ \mid\  u?d.S\ \mid\ S + S\ \mid\ S^{*},
\end{array}
\]
where $c \in \Cact$, $u \in \Uact$, $\ell, k \in \{0, 1\}$, $d \in \Data$, and $E \subseteq \{\comn{f}{m}{n}{} \mid f \in \Chan, \ m, n \in \bbbn\}$.
To implement broadcast communication in the case when the supervisor sends control signals to several distributed components, which do not have to receive the control signals simultaneously, one would also need to introduce action priorities, cf.~\cite{pabookold}. Due to page restrictions, we will not employ broadcast in the general form in this paper and, instead, we enforce three-way communication by employing only the encapsulation operator.

\paragraph{Supervisory coordination of an automated production line}

\begin{figure}[!t]
\centering
\includegraphics[width = 0.7\textwidth]{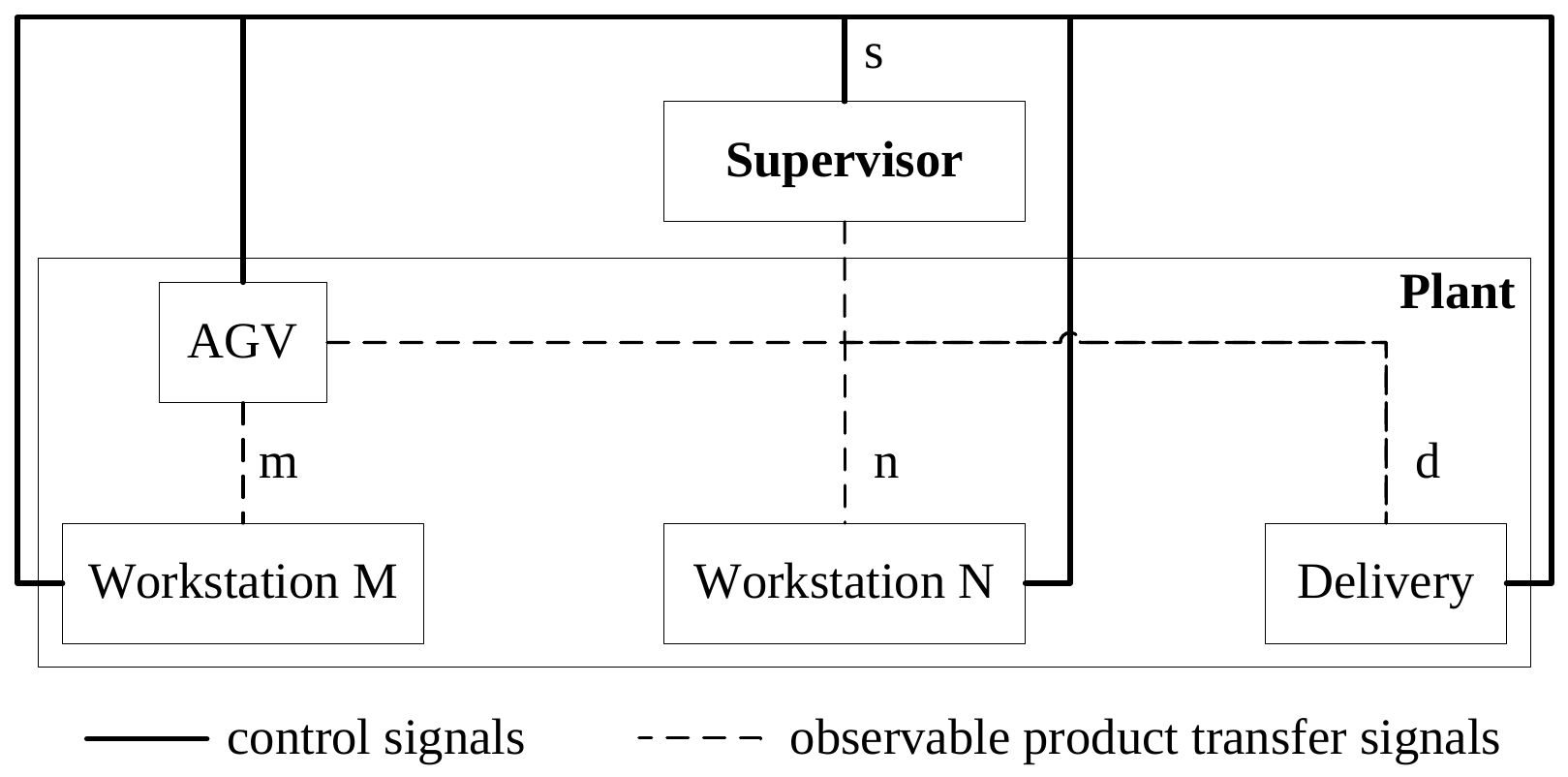}
\caption{Automated production line}\label{fig:agv}
\end{figure}

To illustrate our approach to supervisory control and the model of the control loop, we discuss a simple example concerning coordination of an automated guided vehicle (AGV) in an automated production line, depicted in Figure~\ref{fig:agv}. The AGV is responsible for transferring the preproduct made by Workstation M to Workstation N and transferring the finished product from Workstation N to the Delivery station. The workstations and the AGV are coordinated by a supervisor, which sends the corresponding control signals. We can model the automated production system depicted in~Figure~\ref{fig:agv} employing $\TCPstar$, where $\mathit{M}$, $\mathit{N}$, $\mathit{A}$, and $\mathit{S}$ are process terms that model Workstation M, Workstation N, AGV, and the supervisor. We note that we abstract from the delivery station, depicted by a single event $\mathit{deliver}$, as it does not contribute to any interesting behavior. We retain the communication channel names as depicted in Figure~\ref{fig:agv}, whereas the data elements are $\Data = \{\mathit{make}, \mathit{move2N}, \mathit{preproduct}, \mathit{product}\}$. The uncontrollable channel names are $\Uact = \{m, n, \mathit{produce, process, move, deliver}\}$, whereas $\Cact = \{s\}$ is the set of controllable channel names.
\[
\begin{array}{l}
\mathit{M} \triangleq (s?\mathit{make}.\mathit{produce(preproduct)}.m!\mathit{preproduct}.\emp)^{*} \smallskip \\
\mathit{N} \triangleq (n?\mathit{preproduct}.\mathit{process(preproduct)}.n!\mathit{product}.\emp)^{*} \smallskip \\
\mathit{A} \triangleq (m?\mathit{preproduct}.s?\mathit{move2N}.\mathit{move(preproduct)}.n!\mathit{preproduct}.\emp + n?\mathit{product}.\mathit{deliver(product)}.\emp)^{*} \smallskip \\
\mathit{S} \triangleq (s!\mathit{make}.s!\mathit{move2N}.\emp)^{*}.
\end{array}
\]

Workstation M repeatedly waits for a command from the supervisor to make a preproduct, which is offered to the AGV once it is made. Workstation N waits for a preproduct from the AGV, which is thereafter processed and offered back to the AGV. The AGV can either pick up a preproduct at Workstation M, after which it asks for permission to move the preproduct to Workstation N, or pick up a finished product at Workstation N and deliver it. Now, the unsupervised plant is given by the process
\[
U \triangleq \encap{F}{M \merge N \merge A}, \mathrm{\ where \ } F = \{m?, m!, n?, n!\}.
\]
At this point, we note that we enforce meaningful communication of uncontrollable channels within the plant by encapsulation and this does not restrict the behavior of the unsupervised plant, but only ensures its meaningful behavior. Following the framework outlined above, it can be readily observed that the plant $U \in \mathcal{P}$ follows the outlined syntax.

In this first modeling instance, we assume that the AGV is responsible for delivering the final product and we propose a supervisor as given by the process~$S$. Note that the supervisor $S \in \mathcal{S}$ follows the outlined syntax and it does not make use of any observed information. Supervisor~$S$ repeatedly gives orders to Workstation M for new products to be made, followed by orders to the AGV to transfer the preproduct to Workstation N. Thus, the automated production system is modeled as
\[
\supp{U}{S} \triangleq \encap{E}{S \merge U}, \mathrm{ \ where \ } E = \{s?, s!\},
\]
which enforces communication of control signals and transfer of (pre)products. One can directly check that $S$ is a valid supervisor by establishing that the supervised plant is partially bisimulated by the original plant with respect to the uncontrollable events. To this end, we must employ renaming of events, as the original plant has open communication actions that wait for synchronization with the supervisor. This renaming function $\ren$ traverses the process terms and renames all open communication actions to succeeded communication actions. We note that we overload the name of the renaming function of the process terms and apply it to the communication action names as well. Also, we only specify the communication actions that are actually renamed. The definition of the renaming operation is given by structural induction in Table~\ref{tab:renaming}.

\begin{table}[!t]
\[
\begin{array}{l}
\ren(\dl) = \dl \qquad \ren(\emp) = \emp \qquad \ren(p^*) = \ren(p)^* \qquad \ren(p \diamond q) = \ren(p) \diamond \ren (q)\quad  \mathrm{\ for \ } \diamond \in \{+, \cdot, \merge\}\smallskip \\
\ren(\comn{c}{m}{n}{d}.p) = \ren(\comn{c}{m}{n}{d}).\ren(p) \mathrm{\ for \ } c \in \Chan, \ d \in \Data, \ m, n \in \bbbn  

\end{array}
\]
\caption{Renaming function}\label{tab:renaming}
\end{table}

Now, in order to verify that the supervisor does not disable uncontrollable events, it is sufficient to verify that it holds that
\[
\supp{U}{S} \pbis{\ActC_\Uact} \ren(U), \mathrm{\ where \ } \ren\colon s?d \mapsto \com[s]{d} \mathrm{\ for \ }d \in \Data,
\]
which can be directly checked. We note that there was no restriction imposed on the control requirements, which in this case coincide with the plant and are, therefore, trivially satisfied.

\paragraph{Nonblocking supervision}

Unfortunately, our automated production system has a deadlock. The main reason for the deadlock is that a second preproduct can come too early, before the first product is completely finished and delivered, which is set off by sending a $s!\mathit{make}$ command too early, i.e., before the processed product has left Workstation N. Then, the AGV picks up the preproduct from Workstation M, but it cannot deliver it to Workstation N, as the latter also waits for a finished product to be picked. A trace that leads to deadlock is
\[
\begin{array}{l}
\com[s]{\mathit{make}}\ \mathit{produce(preproduct)}\  \com[m]{\mathit{preproduct}}\ \com[s]{\mathit{move2N}}\ \com[n]{\mathit{preproduct}} \\
\com[s]{\mathit{make}}\ \mathit{produce(preproduct)}\ \com[m]{\mathit{preproduct}}\ \com[s]{\mathit{move2N}}\ \mathit{process(preproduct)}\ \dl.
\end{array}
\]
Such form of blocking behavior appears often, so in many cases the supervisor is additionally required to prevent deadlock and/or livelock, or also known as blocking, behavior~\cite{rwsupervisor,Cassandras}. To this end, special marked states are introduced to automata in supervisory control. We note that these states roughly correspond to successful termination in our setting. The correspondence is not strict, mainly due to the absence of sequential composition and the Kleene star operator in the supervisory control literature and the role of the successful termination in these contexts, confer Table~\ref{tab:TCPsostable}. Note that the marked states do not contribute to the formation of the recognized language of an automaton, which is different from its marked language~\cite{rwsupervisor,Cassandras}.

So, besides the control requirements, we impose an additional deadlock-freedom requirement on the supervisor, stated formally as: there exists no trace $t \in \ActC^*$ such that $\supp{U}{S} \steps{t} \dl$. To ensure this additional nonblocking requirement, we have to modify the supervisor to accept requests for making a new preproduct only after the finished product has been loaded on the AGV, to be transferred to the delivery station. To this end, the supervisor should allow for a new product to be made only after the finished product has been loaded to the AGV at Workstation N, which can be achieved by observing this additional information on channel $n$.

To this end, we modify the supervisor to $S'$ as follows:
\[
S' \triangleq (s!\mathit{make}.s!\mathit{move2N}.n?\mathit{product}.\emp)^{*}.
\]
At this point, we note that communication on the channel $n$ now must occur between three parties, i.e., Workstation N that sends information and the AGV and the supervisor that receive it. In order to enforce this communication, we employ the generic communication actions, i.e., we encapsulate all (incomplete) communication actions on $n$, except for $\comn{n}{1}{2}{\mathit{product}}$. The definition of the deadlock-free supervised plant now becomes:
\[
\supp{U}{S'} \triangleq \encap{E'}{S' \merge U}, \mathrm{\ where\ } E' = \{s?, s!, n?, \com[n]{}\}.
\]
Again, one directly verifies that the supervisor is valid by establishing partial bisimilarity between the supervised and the original plant following an appropriate renaming of the incomplete communication actions, given by $\ren \colon s?d \mapsto \com[s]{d}, \ \com[n]{d} \mapsto \comn{n}{1}{2}{d}$ for $d \in \Data$.

Next, we extend the process theory $\TCPstar$ to accommodate state-based observations as well.

%

\newcommand{\val}[2]{\langle #1, #2 \rangle}
\newcommand{\inacc}{\bot}
\newcommand{\boolf}{\mathcal{B}}
\newcommand{\valu}{\mathcal{V}}
\newcommand{\vt}{\mathrm{T}}
\newcommand{\vf}{\mathrm{F}}
\newcommand{\gc}[2]{#1 :\rightarrow #2}
\newcommand{\cons}[1]{\mathalpha{#1 \searrow\,}}
\newcommand{\TCPstarcons}{\ensuremath{\mathrm{TCP}^{*}_{\mkern-3mu\bot}}}
\newcommand{\inp}[1]{\mathit{in}(#1)}
\newcommand{\stn}[1]{\mathbf{#1}}

\section{Control Loop with State-Based Observations}

We propose $\TCPstarcons$, an extension of $\TCPstar$, with propositional signals~\cite{signalsBaeten} and guarded commands in order to support the modeling of a control loop with state-based observations. To this end, we employ the Boolean algebra
\[\mathbb{B}= (\mathcal{N},\vf,\vt,\neg,\wedge,\vee,\supset),\]
where $\mathcal{N}=\{P_1,\ldots,P_n\}$ are the propositional symbols, the constants represent false and true, whereas the operators denote negation, conjunction, disjunction, and implication, respectively. We use $\boolf$ to denote the standard Boolean expressions of $\mathbb{B}$, which are evaluated with respect to a given valuation $v \colon \boolf \to \{\vf, \vt\}$. The set of valuations is denoted by $\valu$.

\paragraph{Process theory $\TCPstarcons$}

We enrich the syntax of $\TCPstar$ and the set of process terms $\PEXP$ with the \emph{inaccessible process} constant, \emph{guarded commands}, and \emph{signal emission}. The inaccessible process, notation $\inacc$, specifies the process in which there are inconsistencies between the valuation of the propositional variables and the emitted propositional signals. Such a state cannot be reached from any consistent state. The guarded command, notation $\gc{\phi}{p}$, specifies a guard $\phi \in \boolf$ that guards a process $p \in \PEXP$. If the guard is successfully evaluated, the process continues behaving as $p \in \PEXP$ or, else, it deadlocks. The root signal emission process $\rse{\phi}{p}$, emits the propositional signal $\phi \in \boolf$ until the process $p \in \PEXP$ takes an outgoing transition, provided that the propositional signal is consistent with the valuation. To be able to evaluate the propositional expressions, we couple the process terms with valuations, notation $\val{p}{v} \in \PEXP \times \valu$. The dynamics of the valuations, with respect to outgoing labeled transitions, is captured by a predefined valuation effect function, given by $\mathit{effect} \colon \ActC \times \valu \to 2^{\valu}$. With respect to the valuation we have to extend the successful termination predicate to $\final{} \in \PEXP \times \valu$ and the labeled transition relation to $\step{} \in \PEXP \times \valu \times \ActC \times \PEXP \times \valu$. We introduce an additional consistency predicate $\cons{} \in \PEXP \times \valu$ that checks whether the state is consistent. The operational rules in Table~\ref{tab:TCPconsistentsos} give the semantics of the new predicate and the transition relation with respect to the new operators. We note that the operational rules of Table~\ref{tab:TCPsostable} have to be enhanced by decorating the process terms with valuations and additional checks for consistency.

\begin{table}[t!]
\[\small
\hspace{-0.5cm}
\begin{array}{c}
  \osrule{}{\cons{\val{\dl}{v}}} \qquad
  \osrule{}{\cons{\val{\emp}{v}}} \qquad
  \osrule{}{\val{\emp}{v} \downarrow} \qquad
  \osrule{}{\cons{\val{a.p}{v}}} \qquad
  \osrule{\cons{\val{p}{v'}}, \ v' \in \mathit{effect}(a, v)}{\val{a.p}{v} \step{a} \val{p}{v'}} \smallskip \\
  \osrule{\val{p}{v} \step{a} \val{p'}{v'}, \ \cons{\val{q}{v}}}
  {\val{p + q}{v} \step{a} \val{p'}{v'}} \qquad
  \osrule{\cons{\val{p}{v}}, \ \val{q}{v} \step{a} \val{q'}{v'}}
  {\val{p + q}{v} \step{a} \val{q'}{v'}}\qquad
  \osrule{\cons{\val{p}{v}}, \ \val{q}{v}\downarrow}{\val{p + q}{v}\downarrow}   \smallskip \\
  \osrule{\val{p}{v}\downarrow, \ \cons{\val{q}{v}}}{\val{p + q}{v}\downarrow} \qquad
  \osrule{\cons{\val{p}{v}}, \ \cons{\val{q}{v}}}{\cons{\val{p + q}{v}}}
\qquad
  \osrule{\val{p}{v}\downarrow, \ \val{q}{v}\downarrow}
  {\val{p \cdot q}{v}\downarrow} \qquad
  \osrule{\val{p}{v}\downarrow,\ \val{q}{v} \step{a} \val{q'}{v'}}
  {\val{p \cdot q}{v} \step{a} \val{q'}{v'}} \smallskip \\
  \osrule{\val{p}{v} \step{a} \val{p'}{v'}, \ \cons{\val{p' \cdot q}{v'}}}
  {\val{p \cdot q}{v} \step{a} \val{p' \cdot q}{v'}} \qquad
  \osrule{\val{p}{v}\downarrow, \ \cons{\val{q}{v}}}{\cons{\val{p \cdot q}{v}}} \qquad
  \osrule{\cons{\val{p}{v}}, \ \val{p}{v}\not\mkern2mu\downarrow}{\cons{\val{p \cdot q}{v}}}
  \smallskip \\
  \osrule{\cons{\val{p}{v}}}{\val{p^*}{v}\downarrow} \qquad
  \osrule{\cons{\val{p}{v}}}{\cons{\val{p^*}{v}}} \qquad
  \osrule{\val{p}{v} \step{a} \val{p'}{v'}}{\val{p^*}{v} \step{a} \val{p'\cdot p^*}{v'}} \qquad
  \osrule{\val{p}{v} \downarrow, \ \val{q}{v}\downarrow}
  {\val{p \merge q}{v}\downarrow} \qquad
  \osrule{\cons{\val{p}{v}}, \ \cons{\val{q}{v}}}
  {\cons{\val{p \merge q}{v}}} \smallskip \\
  \osrule{\val{p}{v} \step{a} \val{p'}{v'}, \ \cons{\val{q}{v}}, \ \cons{\val{q}{v'}}}
  {\val{p \merge q}{v} \step{a} \val{p' \merge q}{v'}} \qquad
  \osrule{\cons{\val{p}{v}}, \ \cons{\val{p}{v'}}, \ \val{q}{v} \step{a} \val{q'}{v'}}
  {\val{p \merge q}{v} \step{a} \val{p \merge q'}{v'}} \smallskip \\
  \osrule{\val{p}{v} \step{\comn{c}{\ell}{k}{d}} \val{p'}{v'}, \ \val{q}{v} \step{\comn{c}{m}{n}{d}} \val{q'}{v''}, \ \cons{\val{p' \merge q'}{v'''}}, \ v''' \in \mathit{effect}(\comn{c}{\ell + m}{k + n}{d}, v), \ \ell + k > 0, \ m + n >0}
  {\val{p \merge q}{v} \step{\comn{c}{\ell + m}{k + n}{d}} \val{p' \merge q'}{v'''}} \smallskip \\
     \osrule{p \downarrow}{\encap{E}{p} \downarrow}
   \qquad
     \osrule{\cons{p}}{\cons{\encap{E}{p}}}
   \qquad
      \osrule{p \step{a{}} p'
              \quad a \not \in \{\comn{c}{m}{n}{d} \mid \comn{c}{m}{n}{} \in E, \ d \in \Data\}}%
           {\encap{E}{p} \step{a{}}
             \encap{E}{p'}}
\smallskip \\
  \osrule{\val{p}{v} \downarrow, \ v(\phi) = \vt}
  {\val{\gc{\phi}{p}}{v}\downarrow} \qquad
  \osrule{\cons{\val{p}{v}}, \ v(\phi) = \vt}
  {\cons{\val{\gc{\phi}{p}}{v}}} \qquad
  \osrule{v(\phi) = \vf}
  {\cons{\val{\gc{\phi}{p}}{v}}} \qquad
  \osrule{\val{p}{v} \step{a} \val{p'}{v'}, \ v(\phi) = \vt}
  {\val{\gc{\phi}{p}}{v} \step{a} \val{p'}{v'}}   \smallskip \\
  \osrule{\val{p}{v} \downarrow, \ v(\phi) = \vt}
  {\val{\rse{\phi}{p}}{v}\downarrow} \qquad
  \osrule{\cons{\val{p}{v}}, \ v(\phi) = \vt}
  {\cons{\val{\rse{\phi}{p}}{v}}} \qquad
  \osrule{\val{p}{v} \step{a} \val{p'}{v'}, \ v(\phi) = \vt}
  {\val{\rse{\phi}{p}}{v} \step{a} \val{p'}{v'}}   \smallskip \\
\end{array}
\]
\caption{Operational rules for \TCPstarcons}
\label{tab:TCPconsistentsos}
\end{table}

The rules ensure that when taking an action transition, the target state is always consistent. We comment the important rules that are not directly taken from Table~\ref{tab:TCPsostable} and adapted in a setting with valuations. The deadlock, successful termination, and action prefix are always consistent as stated by rules~18, 19, and 21, respectively. The target process must be consistent for the target valuation, which is determined by the effect function as given by rule 22. Rules 23-27 introduce valuations and consistency for the alternative composition, whereas rules 28-32 do the same for the sequential composition and rules 33-35 describe iteration. Rules 38 and 39 introduce interleaving in the new setting. Rule 40 shows how the effect function is impacted by synchronization. For the effect function to be well-defined with respect to the valuations by interleaving and synchronization~\cite{pabook}, we require additionally that
\[
\begin{array}{l}
\mathit{effect}(\comn{c}{\ell + m}{k + n}{d}, v) \subseteq \mathit{effect}(\comn{c}{m}{n}{d}, \mathit{effect}(\comn{c}{\ell}{k}{d}, v)) \cap 
\mathit{effect}(\comn{c}{\ell}{k}{d}, \mathit{effect}(\comn{c}{m}{n}{d}, v))
\end{array}
\]
for all $\ell, k, m, n \in \bbbn$ with $\ell + k >0$ and $m + n > 0$. Rules 41-43 introduce the encapsulation operator in the new setting. Rules 44 and 47 show that a guarded process does not deadlock only when the guard evaluates to true. We note, however, that the value of the guard does not affect the consistency of the term, provided that the term that is guarded is consistent. This is in direct contrast with signal emission, see rule 49, where the consistency is preserved only if the emitting signal is consistent within the valuation. In that case, the process that emits the signal can continue with its normal execution.

Finally, we also have to adapt our behavioral relation in order to correctly handle the valuations. Here, we directly employ the approach of~\cite{signalsBaeten,pabook}, where this extension is shown for bisimulation. We consider a relation~$R \subseteq \PEXP \times \PEXP$ to be a partial bisimulation with respect to a bisimulation action set~$B \subseteq \ActC$, if for all $(p,q) \in R$ it holds that:
\begin{enumerate}
\item  if $\val{p}{v}\downarrow$ for some $v \in \mathcal{V}$, then $\val{q}{v}\downarrow$;

\item 
if $\val{p}{v} \step{a} \val{p'}{v'}$ for some $v \in \mathcal{V}$ and $a \in \ActC$, then there exists $q' \in \PEXP$ such that $\val{q}{v} \step{a} \val{q'}{v'}$ and $(p',q') \in R$;

\item 
if $\val{q}{v} \step{b} \val{q'}{v'}$ for some $v \in \mathcal{V}$ and $b \in B$, then there exists $p \in \PEXP$ such that $\val{p}{v} \step{b} \val{p'}{v'}$ and $(p',q') \in R$.
\end{enumerate}
Again, if $(p, q) \in R$, we say that $p$ is partially bisimilar to $q$ with respect to $B$ and we write $p \pbis{B} q$. If $q \pbis{B} p$ holds as well, we write $p \bpbis{B} q$. Also, we consider a process $s \in \PEXP$ to be a supervisor of the plant $p \in \PEXP$ with respect to the control requirements $r \in \PEXP$ if $\supp{p}{s} \pbis{\emptyset} r$ and $\supp{p}{s} \pbis{\ActC_\Uact} p$.

\paragraph{Plant and supervisor syntax} Now, we can model the control loop with state-based observations as depicted in Figure~\ref{fig:control-loop}c). Intuitively, the plant emits a signal that identifies the observable states. Upon observing such a signal, the supervisor checks which controllable actions are allowed in the state identified by the signal. Allowance of actions is specified in the form of guarded prefixes in which a process term is bound to a propositional formula deduced from the control requirements. These new concepts introduce further asymmetry in the control loop, where the syntax of the plant and the supervisor is again given by $P$ and $S$, respectively:
\[
\begin{array}{l}
P ::= \dl\,\mid\,\emp\,\mid\,c?d.P\,\mid\, \comn{u}{\ell}{k}d.P\,\mid\, P \cdot P \,
  \mid\,\,P + P\, \mid\, P \merge P\,\mid\,\encap{E}{P}\,\mid\,
  \gc{\phi}{P}\,\mid\,
  \rse{\phi}{P} \,\mid\,  P^{*}
\smallskip \\
S ::= \emp\ \mid c!d.S\ \mid\ S + S\ \mid\ \gc{\phi}{S}\ \mid\ S^{*},
\end{array}
\]
for $c \in \Cact$, $u \in \Uact$, $\ell, k \in \{0, 1\}$, $d \in \Data$, $\phi \in \boolf$, and $E \subseteq \{\comn{f}{m}{n}{} \mid f \in \Chan, \, m, n \in \bbbn\}$.

We note that in the state-based setting, the control requirements can be stated directly in terms of states, i.e., signals that the state is emitting, and additionally, one can specify which events are allowed with respect to the emitted signals. The control requirements $\mathcal{R}$ have the following syntax given by $R$:
\[
R ::= \  \phi \  \mid \ \step{\comn{f}{m}{n}{d}} \supset \phi \ \mid\  \phi \supset \nostep{\comn{f}{m}{n}{d}},
\]
for $f \in \Chan$, $d \in \Data$, $m, n \in \bbbn$, and $\phi \in \boolf$. Given control requirements $r \in \mathcal{R}$ are satisfied with respect to process $p \in \PEXP$ in the  valuation $v \in \valu$, notation $\val{p}{v} \models r$, according to the following operational rules:
\[
\osrule{v(\phi) = \vt}{\val{p}{v} \models \phi} \qquad
\osrule{\val{p}{v} \models \neg \phi \supset \nostep{\comn{f}{m}{n}{d}}}{\val{p}{v} \models \step{\comn{f}{m}{n}{d}} \supset \phi} \qquad
\osrule{v(\phi) = \vt, \ \val{p}{v} \nostep{\comn{f}{m}{n}{d}} {}}{\val{p}{v} \models \phi \supset \nostep{\comn{f}{m}{n}{d}}},
\]
where $\val{p}{v} \nostep{a} {}$ for $a \in \ActC$ holds if there does not exist $\val{p'}{v'}$ such that $\val{p}{v} \step{a} \val{p'}{v'}$ with $v' = \mathit{effect}(a, v)$. We note that the second form of the control requirements is introduced since it corresponds better to modeling intuition and it is equivalent to the third, which is easily seen from the operational rule~52. Furthermore, for the propositional symbols, we employ the notation $\inp{\stn{StateName}}$, where $\inp{\stn{StateName}}$ is a signal emitted from the process, corresponding to a state in the labeled graph representation identified by $\stn{StateName}$. For example, in the Current Power Mode process in Figure~\ref{fig:caseautomata}, the process modeling the state with associated name $\stn{Standby}$ emits the signal $\inp{\stn{Standby}}$.

\newcommand{\vev}[1]{\mathit{#1}}

\section{Coordination Control of Maintenance Procedures}

\begin{figure}[!t]
  \centering
  \includegraphics[width=\columnwidth]{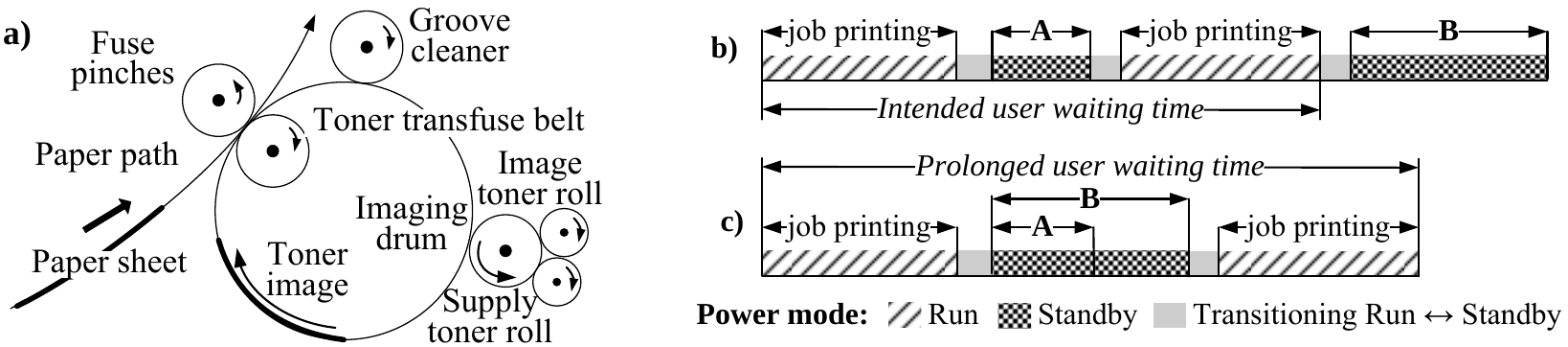}\\
  \caption{a) Printing process, b) maintenance operation, c) emergent behavior}\label{fig:printer}
\end{figure}

We employ the process theory $\TCPstarcons$ to model the coordination of maintenance procedures of a printing process of a high-end Oc\'{e} printer~\cite{wodes2010}. The printing process consists of several distributed independent components as depicted in Figure~\ref{fig:printer}a). The process applies the toner image onto the toner transfuse belt and fuses it onto the paper sheet. To maintain high printing quality, several maintenance operations have to be carried out, like: toner transfuse belt jittering, which displaces the transfuse belt to prolong its lifespan due to wearing by paper edges; black image operation, which removes paper dust by occasionally printing completely black pages; coarse toner particles removal operation; etc. Most maintenance operations are scheduled after a given number of prints, but must be carried out after a given strict threshold.
To perform a maintenance operation, the printing process has to change its power mode, from Run mode, used for printing, to Standby mode, required for maintenance. However, this change can actually trigger pending maintenance operations, which may unnecessary prolong the user waiting time.

As an illustration, in Figure~\ref{fig:printer}b) we depict the situation, where due to inevitable execution of maintenance operation \textbf{A}, the ongoing print job is suspended and the power mode of the printer is changed to Standby. However, an unwanted situation occurs, i.e., the power mode change triggers a longer, yet postponable maintenance operation \textbf{B} as depicted in Figure~\ref{fig:printer}c). For instance, a black image operation (\textbf{A}) must be performed, which takes the time needed to print one page and is activated often, but the switching of the power mode triggers the much longer toner transfuse belt jittering (\textbf{B}), thus making the user wait unnecessarily.

The goal of the research performed for this use case was to eliminate
undesired emergent behavior due to interactions of otherwise
correctly-functioning distributed components, with primary focus at
coordinating maintenance operations. Our approach was to synthesize a
supervisory coordinator for the maintenance procedures~\cite{wodes2010}, which here we model in the proposed process theory.

\paragraph{Informal description of the printing process}

\begin{figure}[!ht]
  \centering
  \includegraphics[scale = 0.6]{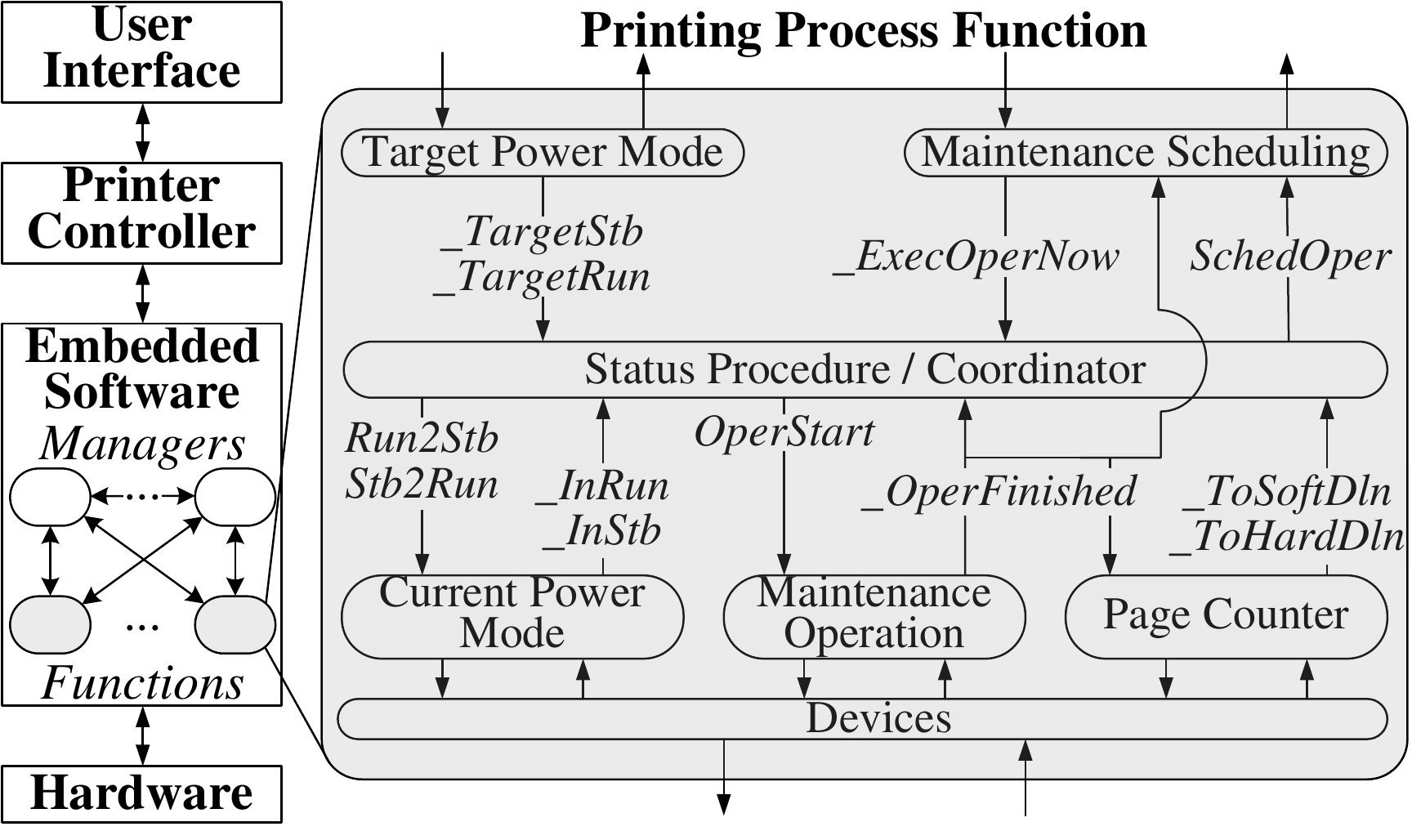}\\
  \caption{Printing process function.}\label{fig:detailedfunction}
\end{figure}

An abstract view of the control architecture of a high-end printer is depicted in Figure~\ref{fig:detailedfunction}. Print jobs are sent to the printer by means of the user interface. The printer controller communicates with the user and assigns print jobs to the embedded software, which actuates the hardware to realize print jobs. The embedded software is organized in a distributed way, per functional aspect, such as, paper path, printing process, etc. Several managers communicate with the printer controller and each other to assign tasks to functions, which take care of the functional aspects.

We depict a printing process function comprising one maintenance operation in Figure~\ref{fig:detailedfunction}. We abstract from all timing behavior, which can be present in some control signals, e.g., execute a maintenance procedure after a given delay. Each function is hierarchically organized as follows: (1) controllers: Target Power Mode and Maintenance Scheduling, which receive control and scheduling tasks from the managers; (2) procedures: Status Procedure, Current Power Mode, Maintenance Operation, and Page Counter, which handle specific tasks and actuate devices, and (3) devices as hardware interface.

The Status Procedure is responsible for coordinating the other procedures given the input from the controllers. It will be implemented as a supervisory coordinator with respect to the coordination rules given below. The Current Power Mode procedure sets the power mode to Run or Standby depending on the enabling signals from the Status Procedure $\vev{Stb2Run}$ and $\vev{Run2Stb}$, respectively. The confirmation is sent back via the signals $\vev{\_InRun}$ and $\vev{\_InStb}$, respectively. Maintenance Operation either carries out maintenance operation or it is idle. The triggering signal is $\vev{OperStart}$ and the confirmation is sent back by $\vev{\_OperFinished}$. The Page Counter procedure counts the printed pages since the last maintenance and sends signals when soft and hard deadlines are reached using $\vev{\_ToSoftDln}$ and $\vev{\_ToHardDln}$, respectively. The counter is reset each time the maintenance is finished, by receiving the confirmation signal $\vev{\_OperFinished}$ from Maintenance Operation. The controller Target Power Mode defines which mode is requested by the manager by sending the control signals $\vev{\_TargetStb}$ and $\vev{\_TargetRun}$ to the Status Procedure. Maintenance Scheduling receives a request for maintenance from Status Procedure via the signal $\vev{SchedOper}$, which it forwards to a manager. The manager confirms the scheduling with the other functions and sends a response back to the Status Procedure via the control signal $\vev{\_ExecOperNow}$. It also receives feedback from Maintenance Operation that the maintenance is finished in order to reset the scheduling.

\paragraph{Plant modeling in $\TCPstarcons$}
\begin{figure*}[!t]
  \centering
  \includegraphics[width=\textwidth]{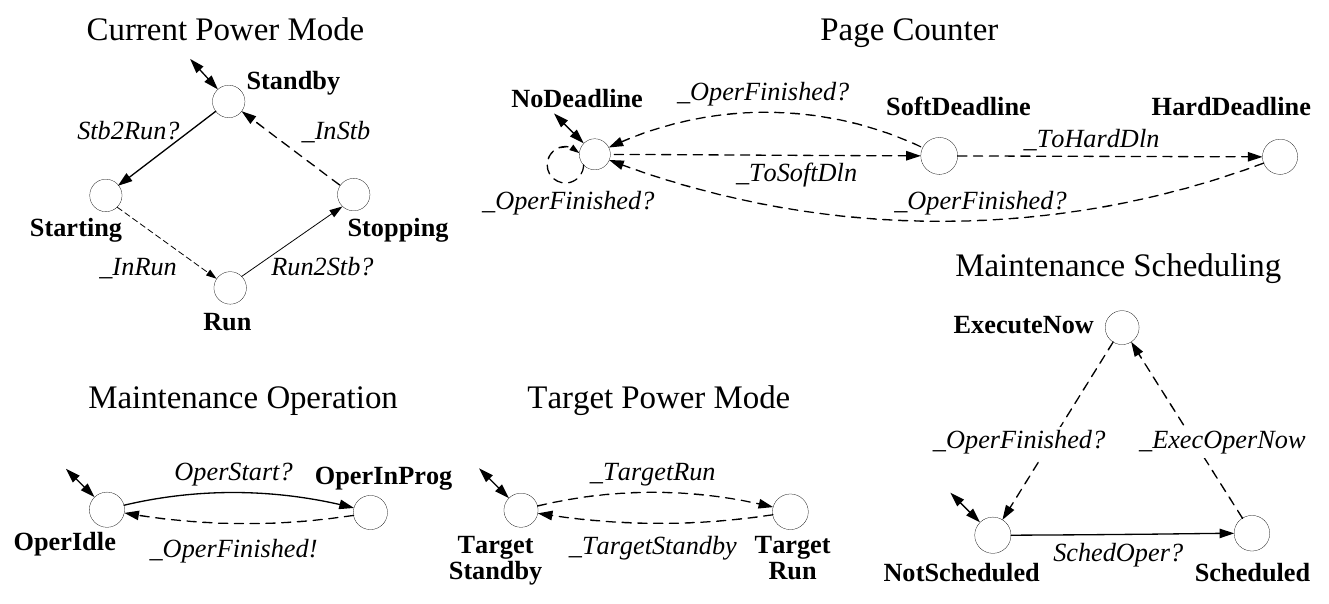}\\
  \caption{Plant modeling of the Printing Process Function.}\label{fig:caseautomata}
\end{figure*}
We model the procedures by means of processes. We retain the names of the control signals, turning them into communication actions where appropriate. The controllable communicating channels are the given by $\Cact = \{\vev{Run2Stb}, \vev{Stb2Run}, \vev{SchedOper},  \vev{OperStart}\}$, modeled as receive actions in the plant. We note that we abstract from data elements as communication should only enforce ordering of events. The other actions are uncontrollable, also prefixed by $\_$, where only $\vev{\_OperFinished}$ is modeled as a communication action, as the procedure Maintenance operation must send signals and reset Page Counter and Maintenance Scheduling. The signals emitted from the plant uniquely identify the state of the plant. For clarity, we also depict the processes in Figure~\ref{fig:caseautomata}, where the signal names are given next to the states that emit them. Page Counter is modeled by the process $C$, where $\vev{\_OperFinished}$ is modeled as a receive action, to be synchronized with Maintenance Operation:\linebreak[4]
\[
\begin{array}{lll}
C & \triangleq & \Big(\rse{\mathit{in}(\textbf{NoDeadline})}{} (\\
&&\qquad \mathit{\_OperFinished?}.1 + {} \\
&&\qquad \mathit{\_ToSoftDln}.\big(\rse{\mathit{in}(\textbf{SoftDeadline})}{} ( \\
&& \qquad \qquad \mathit{\_OperFinished?}.1+ \mathit{\_ToHardDln}.\rse{\mathit{in}(\textbf{HardDeadline})}{\mathit{\_OperFinished?}.1}))\big)\Big)^*.
\end{array}
\]

Maintenance Operation is specified by the process $O$, where $\vev{\_OperFinished}$ broadcasts that the maintenance operation has finished:
\[
\begin{array}{lll}
 O &\triangleq &\big(\rse{\mathit{in}(\textbf{OperIdle})}{\mathit{OperStart?}.
  \rse{\mathit{in}(\textbf{OperInProg})}{\mathit{\_OperFinished!}.1}}\big)^*.
\end{array}
\]

Target Power Mode is modeled by $T$:
\[
\begin{array}{lll}
T&\triangleq&\big(\rse{\mathit{in}(\textbf{TargetStandby})}{\mathit{\_TargetRun}}. \rse{\mathit{in}(\textbf{TargetRun})}{\mathit{\_TargetStandby}.1}\big)^*,
\end{array}
\]
whereas Current Power Mode is given by $P$:
\[
\begin{array}{lll}
P &\triangleq &\big(\rse{\mathit{in}{(\textbf{Standby})}}{\mathit{Stb2Run?}.\rse{\mathit{in}{(\textbf{Starting})}}{\mathit{\_InRun}}}.
\\
  && \phantom{(}
\rse{\mathit{in}(\textbf{Run})}{
  \mathit{Run2Stb?}.\rse{\mathit{in}(\textbf{Stopping})}{\mathit{\_InStb}}}.1\big)^*.
\end{array}
\]

Finally, Maintenance Scheduling is specified as $M$:
\[
\begin{array}{lll}
M & \triangleq & \big(\rse{\mathit{in}(\textbf{NotScheduled})}{\mathit{SchedOper?}}.\rse{\mathit{in}(\textbf{Scheduled})}{\mathit{\_ExecOperNow}}. \\
  && \phantom{(}
     \rse{\mathit{in}(\textbf{ExecuteNow})}{\mathit{\_OperFinished?}.1}\big)^*.
\end{array}
\]
Due to the generic valuation effect function, we need to impose additional restriction on the emitted signals. More precisely, we wish that the signals emitted in a process are not ambiguous, e.g., it cannot be that both $\inp{\stn{Standby}}$ and $\inp{\stn{Run}}$ are valid at the same time as these are two distinct states that belong to the same process. Note, however, that this situation is possible as one can easily construct a valuation effect function that always assigns the same values to the above propositional symbols. However, such misconstrued valuations can actually lead to wrong supervised behavior as the supervisor bases its decision on the emitted signals, which are deduced from the valuations. At this point, we have two viable options. One is to make the signal emission complete and rewrite all signal emissions such that the effect function leads to inconsistencies unless it uniquely defines each state. For example, then we would have to rewrite $T$ to $T'$:
\[
\begin{array}{lll}
T'&\triangleq&\big(\rse{(\inp{\stn{TargetStandby}} \wedge \neg \inp{\stn{TargetRun}})}{\mathit{\_TargetRun}}. \\
&&\phantom{\big(}\rse{(\neg \inp{\stn{TargetStandby}} \wedge \inp{\stn{TargetRun}})}{\mathit{\_TargetStandby}.1}\big)^*,
\end{array}
\]
and adapt the rest of the processes analogously. The other option is to set an invariant process in parallel to the components that will ensure that only one state can be identified per process. To this end, we define the operation $\textstyle \bigoplus_{P \in S} P \triangleq \bigvee_{P \in S} \left(P \wedge \bigwedge_{Q \in S \setminus \{P\}} \neg Q\right)$ for a set of propositional symbols $S \subseteq \mathcal{N}$, which ensures that only one propositional symbol, i.e., one signal, is exclusively emitted per state. Now, the invariant process $I$ that enforces this restriction can be specified as:
\[
\begin{array}{lll}
I & \triangleq & \left( \rse{\left(\bigwedge_{i = 1}^5 \bigoplus_{P \in \{S_i\}}P\right)}\dl\right)^*,
\end{array}
\]
where $S_i \subset \mathcal{N}$ for $i \in \{1, \ldots, 5\}$ contain the signals emitted by the processes $C$, $O$, $T$, $P$, and $M$, respectively, i.e.,
\[
\begin{array}{l}
S_1 = \{\inp{\stn{NoDeadline}}, \inp{\stn{SoftDeadline}}, \inp{\stn{HardDeadline}}\}, \\
S_2 = \{\inp{\stn{OperIdle}}, \inp{\stn{OperInProg}}\},\\ 
S_3 = \{\inp{\stn{TargetStandby}}, \inp{\stn{TargetRun}}\},\\ 
S_4 = \{\inp{\stn{Standby}}, \inp{\stn{Starting}}, \inp{\stn{Stopping}}, \inp{\stn{Run}}\},\\ S_5 = \{\inp{\stn{NotScheduled}}, \inp{\stn{Scheduled}}, \inp{\stn{ExecuteNow}}\}.
\end{array}
\]
Finally, the unsupervised plant can be specified as $U \in \mathcal{P}$ given by:
\[
U \triangleq \encap{F}{C \merge O \merge T \merge P \merge M} \merge I,
\]
where $F = \{\_OperFinished?, \_OperFinished!, \comn{\_OperFinished}{0}{2}, \com[\_OperFinished]{}\}$ enforces a three-way communication between $C$, $O$, and $M$. We note that due to the stringent streamlining invariant, the role of the valuation effect function is now diminished and one can simply assume that $\mathit{effect}(a, v) = \mathcal{V}$ for every $a \in \ActC$ and $v \in \mathcal{V}$.

\paragraph{Coordination requirements}

We synthesized a coordinator that implements Status Procedure, see Figure~\ref{fig:detailedfunction}, which coordinates the maintenance procedures with the rest of the printing process. The following coordination requests describe the behavior of the Status Procedure:
\begin{enumerate}
\item Maintenance operations can be performed only when the printing process is in standby;

\item Maintenance operations can be scheduled only if soft deadline has been reached and there are no print jobs in progress or a hard deadline is passed;

\item Maintenance operations can be started only after being scheduled;

\item The power mode of the printing process must follow the power mode dictated by the managers, unless overridden by a pending maintenance operation.
\end{enumerate}

We formalize these control requirements as follows:

\begin{enumerate}
\item The maintenance procedure is performed if the process emits the signal $\inp{\stn{OperInProg}}$, while emitting the signal $\inp{\stn{Standby}}$ as well:
      \[
        R_1 \triangleq \inp{\stn{OperInProg}} \supset \inp{\stn{Standby}}.
      \]
\item For the control signal \emph{SchedOper!} to be sent to Maintenance
      Scheduling, either one of the following must hold: (1) A soft deadline has been passed, identified by emission of the signal $\inp{\stn{SoftDeadline}}$, and there are no print jobs waiting, meaning that the target power mode is not in run, identified by the signal $\inp{\stn{TargetRun}}$; or (2) A hard deadline has been passed, indicated by the signal $\inp{\stn{HardDeadline}}$. This is captured by the following control requirement:
      \[
      \begin{array}{l}
      R_2 \triangleq \step{\vev{SchedOper!}} \supset (\inp{\stn{SoftDeadline}} \wedge\neg \inp{\stn{TargetRun}}) \vee  \inp{\stn{HardDeadline}}.
      \end{array}
      \]
\item The maintenance operation can be started by sending the control signal \emph{OperStart!} only if it has been scheduled, prompted by the emission of the signal $\inp{\stn{ExecuteNow}}$:
      \[
        R_3 \triangleq \step{\vev{OperStart!}} \supset \inp{\stn{ExecuteNow}}.
      \]
\item If we want to switch from standby to run power mode, indicated by sending the control signal \emph{Stb2Run!}, then this has been requested by the target power mode manager by emitting the signal $\inp{\stn{TargetRun}}$, provided that there are no maintenance operations scheduled, for which the signal $\inp{\stn{ExecuteNow}}$ should be checked:
      \[
        R_{4,1} \triangleq \step{\vev{Stb2Run}} \supset \inp{\stn{TargetRun}}
        \wedge\neg \inp{\stn{ExecuteNow}}.
      \]
When switching from run to standby power mode, indicated by sending the control signal \emph{Run2Stb!}, the target power mode should be in standby, given by emission of the signal $\inp{\stn{TargetStandby}}$. An exception is made when a maintenance
      operation is scheduled to be executed, given by emission of the signal $\inp{\stn{ExecuteNow}}$:
      \[
        R_{4,2} \triangleq \step{\vev{Run2Stb}} \supset \inp{\stn{TargetStandby}} \vee \inp{\stn{ExecuteNow}}.
      \]
\end{enumerate}

\paragraph{Supervisor synthesis}

With respect to the control requirements we synthesized a deadlock- and livelock-free maximally-permissive supervisor~\cite{wodes2010}. The supervisor sends the control signals upon observation of certain signal combinations, which are given in the form of guards. The indices of the guards correspond to the indices of the control requirements that concern the control signal:
\[
\begin{array}{l}
g_2 \triangleq (\mathit{in}(\textbf{SoftDeadline})\wedge\mathit{in}(\textbf{TargetStandby}))\vee
  \mathit{in}(\textbf{HardDeadline}) \smallskip \\
g_3 \triangleq \mathit{in}(\textbf{Standby})\wedge\mathit{in}(\textbf{ExecuteNow}) \smallskip \\
g_{4,1} \triangleq \neg\mathit{in}(\textbf{ExecuteNow})\wedge\mathit{in}(\textbf{TargetRun})
  \wedge\neg\mathit{in}(\textbf{OperInProg}) \smallskip \\
g_{4,2} \triangleq (\neg\mathit{in}(\textbf{ExecuteNow})\wedge\mathit{in}(\textbf{TargetStandby}))\vee
  \mathit{in}(\textbf{ExecuteNow}).
\end{array}
\]
The supervisor is given by $S \in \mathcal{S}$:
\[
\begin{array}{l}
S \triangleq \Big( \gc{g_2}{\vev{SchedOper!}}.\emp + \gc{g_3}{\vev{OperStart!}}.\emp + {} \gc{g_{4,1}}{\vev{Stb2Run!}}. \emp + \gc{g_{4,2}}{\vev{Run2Stb!}}.\emp \Big)^*.
\end{array}
\]
Now, the supervised plant $\supp{U}{S}$ is given by:
\[
\supp{U}{S} \triangleq \encap{E}{S \merge U}, \mathrm{\ where \ } E = \{c!, c? \mid c \in \Cact\}.
\]
Again, we can show that the supervised plant is partially bisimilar to the original plant with respect to the uncontrollable events by showing that
\[
\supp{U}{S} \pbis{\ActC_\Uact} \ren(U), \mathrm{\ where \ } \ren\colon c? \mapsto \com[c]{} \mathrm{\ for \ }c \in \Cact.
\]

\begin{figure}[!t]
\centering
\includegraphics[width = 0.8 \textwidth]{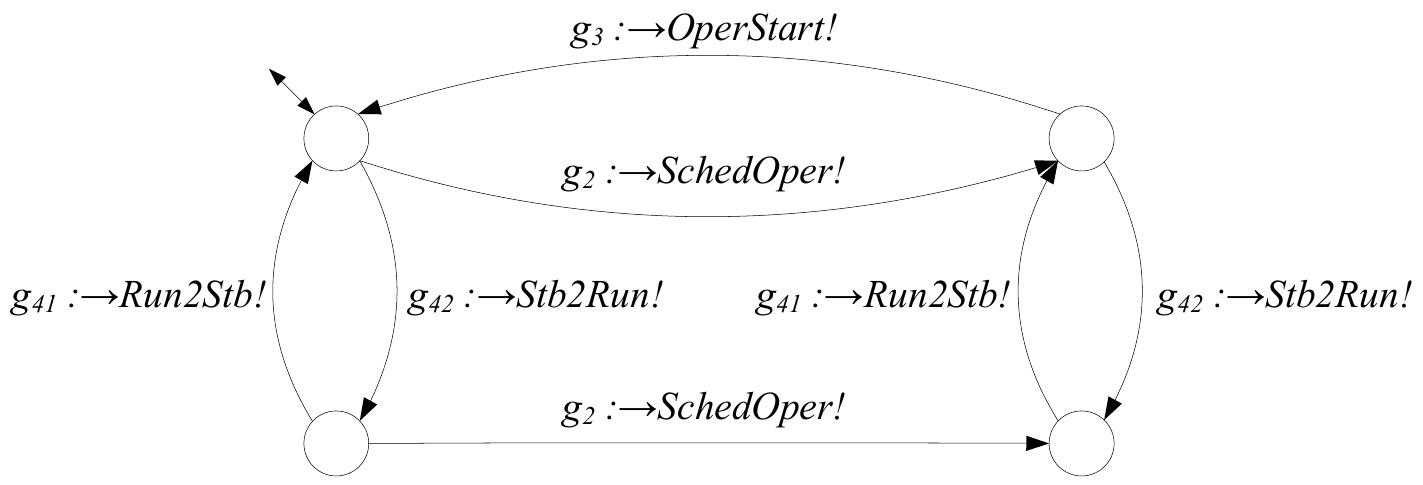}
\caption{Alternative form of the supervisor}\label{fig:supervisor}
\end{figure}

The above form of the supervisor does not provide much information regarding the choices made. It can be visualized as a single state transition system with four outgoing guarded transitions. However, it is not difficult to deduce that initially the event \emph{Run2Stb} is not possible since the initial signal is $\inp{\stn{Standby}}$. Also, \emph{StartOper} is initially unavailable as the signal $\inp{\stn{ExecuteNow}}$ is not emitted. In order to better understand the consequences of the control choices made by the supervisor and the thereafter enabled controllable events, we depict an alternative supervisor in Figure~\ref{fig:supervisor}. We note that both variants of the supervisor produce equivalent supervised behavior (the guards remain the same), the difference being that the supervisor depicted in Figure~\ref{fig:supervisor} reveals the consequences of choosing a particular controllable action. We can now observe, that if the operation is scheduled while the printing process is in standby power mode, then it can be directly executed, returning the supervisor to the initial state. However, if the power mode is run, then the maintenance operation can still be scheduled, but the system has to switch to standby power mode before it can be executed.

\section{Conclusions and Future Work}

We modeled two prominent types of supervisory control loops, one employing event-based observations and the other employing state-based observations. To this end, we revisited the process theory $\TCPstar$ of~\cite{pabook}, where we introduced generic communication actions to model communication between multiple parties, and we applied the developed theory to model the control loop with event-based observations. We classified the processes modeling the unsupervised system and the controller to capture their specific goals. We illustrated our approach on an academic example of coordinating an automated guided vehicle in a production line. To model the control loop with state-based observations as well, we extended the process theory with guarded commands and root signal emission, leading to $\TCPstarcons$. We reiterated on an industrial study dealing with coordination of maintenance procedures in a printing process of a high-tech printer. We demonstrated that our approach is capable of modeling the interaction in the control loop precisely by distinguishing between the information flows of the observations and the control signals.

\paragraph{Application of process theory in supervisory coordination}

The work presented in this paper is merely the third step in our investigations regarding application of process theory in supervisory control and coordination. Our prior work identified and employed partial bisimulation as a suitable behavioral relation to capture the central notion of controllability~\cite{acc2011}. Based on this relation we developed an efficient minimization procedure for nondeterministic plants that respects controllability. Here, we modeled the most prominent variants of the supervisory control loop and further calibrating the process algebra with respect to the notions that are needed to correctly capture the central notions of supervisory control theory.

The issues are far from resolved. We intend to proceed in several directions of research, where we expect that a process-theoretic approach can advance the theory and/or define the notion more clearly and concisely. One issue that we partially treat in this paper is the notion of partial observability, which is an inherent property of plants in which due to unavailability of sensors certain information is unobservable to the supervisor~\cite{Cassandras}. There is a lot of work regarding partial observability of events, which can be treated as uncontrollable actions that are not communicated to the supervisor or as silent steps from which the supervisor has to abstract. The first option is already present in the current setting, whereas the second approach is more than familiar in the process-theoretic community. An unavoidable complication in supervisory control is that the supervisor must not make a wrong control choice, irrespective of not being able to observe the correct state of the plant, making partial observability a global property~\cite{acc2011}. In the setting with state-based observations, one can easily abstract from state information by emitting slightly ambiguous signals, e.g., instead of uniquely identifying as being in states $\stn{S}$ or $\stn{T}$, one can emit the signal $\inp{\stn{S}} \vee \inp{\stn{T}}$. We intend to further investigate the mechanics of state abstraction in supervisory control.

As expected, there are quantitative extensions of supervisory control theory employing real and stochastic timing, probabilities, and data. However, the supervisory control community seems to struggle with clear and acceptable definitions of controllability, as typically these follow the original approach of~\cite{rwsupervisor} and are, thus, given in trace semantics. There are other approaches that are instead based on games, but these often suffer from great computational complexities. We believe that here the community of process theory and verification can contribute a great deal, both in providing suitable definitions and algorithms for minimization and supervisor synthesis. Finally, the supervisor synthesis algorithms almost always have distributed, decentralized, modular, or hierarchical implementations. Concurrency is inherent to our work, and we believe that there are a lot of interesting problems, issues, and challenges that are hidden in this exciting field.

%


\bibliographystyle{eptcs}
\bibliography{bibliography}


\end{document}